\documentclass{emulateapj}
\usepackage{apjfonts,amsmath}

\begin{document}
\title{An upper limit on the mass of the black hole in 
Ursa Minor dwarf galaxy}
\author{V.~Lora\altaffilmark{1}, F.~J.~S{\'a}nchez-Salcedo\altaffilmark{1},
 A.~C.~Raga\altaffilmark{2} and A.~Esquivel\altaffilmark{2}}
\altaffiltext{1}{Instituto de Astronom\'ia, Universidad Nacional Aut\'onoma
de M\'exico, Ciudad Universitaria, Ap.~70-468, C.P.~04510 Mexico City, Mexico;
vlora@astroscu.unam.mx, jsanchez@astroscu.unam.mx}
\altaffiltext{2}{Instituto de Ciencias Nucleares, Universidad Nacional Aut\'onoma
de M\'exico, Ciudad Universitaria, Ap.~70-543, C.P.~04510 Mexico City, Mexico;
raga@nucleares.unam.mx,esquivel@nucleares.unam.mx}
\begin{abstract}
The well-established correlations between the mass of massive black holes (BHs)
in the nuclei of most studied galaxies and various global properties of their
hosting galaxy lend support to the idea that dwarf galaxies and
globular clusters could also host a BH in their centers. 
Direct kinematic detection of BHs in dwarf spheroidal (dSph) galaxies 
are seriously hindered by the small number of stars inside the
gravitational influence region of the BH.
The aim of this Letter is to establish an upper dynamical limit on the mass 
of the putative BH in the Ursa Minor (UMi) dSph galaxy.
We present direct N-body simulations of the tidal disruption of 
the dynamical fossil observed in UMi, with and without a massive BH.
We find that the observed substructure is incompatible with the presence of a 
massive BH of $(2-3)\times 10^{4}$ $M_{\odot}$ within the
core of UMi.
These limits are consistent with the extrapolation of the $M_{BH}-\sigma$ 
relation to the $M_{BH}<10^{6}$ $M_{\odot}$ regime.
We also show that the BH may be off-center with respect to the center 
of symmetry of the whole galaxy.

\end{abstract}


\maketitle

\section{INTRODUCTION}

Intermediate-mass black holes (IMBH; $M_{BH} \sim 10^{2}-10^{4}M_{\odot}$)
accreting gas from their surroundings have been postulated to explain
the engines behind ultraluminous X-ray sources recently discovered in
nearby galaxies (see Colbert \& Miller 2005, for a review).
IMBHs would fill the existing gap between stellar 
and supermassive black holes found in active galactic nuclei. 
According to the tight relation between $M_{BH}$ and the
central velocity dispersion $\sigma$ (Gebhardt et al.~2000; Ferrarese
\& Merrit 2000; Tremaine et al.~2002), or between $M_{BH}$ and the
mass of the bulge (Magorrian et al.~1998), 
a natural place to look for these IMBHs are astrophysical systems less 
massive than normal
galaxies, such as dense star clusters, globular clusters, and dwarf galaxies. 
If $M_{BH}$ is correlated with the total gravitational mass of their
host galaxy (Ferrarese 2002; Baes et al.~2003), 
central BHs should be an essential
element in dark matter dominated objects -such as dwarf spheroidal (dSph)
galaxies. 
Estimates of the mass of BH in these systems would be of great importance
in completing the $M_{BH}-\sigma$ relation.

Direct and indirect searches for IMBH at the centers of globular 
clusters and small galaxies 
have been attempted (e.g., Gerssen et al.~2002, 2003; Valluri et al.~2005; 
Maccarone et al.~2005; Ghosh et al.~2006; Maccarone et al.~2007;
Noyola et al.~2008).
For instance, Noyola et al.~(2008) have reported a central density
cusp and higher velocity dispersions in their central field of the 
globular cluster $\omega$ Centauri, which could be due to a central BH 
($M_{BH}\sim 4\times 10^{4}M_{\odot}$). 
However, a new analysis by Anderson \& van der Marel (2009) and
van der Marel \& Anderson (2009) does not confirm the arguments given
by Noyola et al.~(2008), and provides an upper limit for the BH mass
of $1.2\times 10^{4} M_{\odot}$.

There is sparse evidence that BHs could be present in at least some dSph.
The Ursa Minor (UMi) dSph galaxy
has been also suspected to contain a BH of $\sim 10^{6}$ $M_{\odot}$
(Strobel \& Lake 1994; Demers et al.~1995). 
Maccarone et al.~(2005) discuss the possibility that the 
radio source found near the core of UMi is, in fact, a BH with a 
mass $\sim 10^{4}M_{\odot}$. 

In this Letter, we examine the dynamical effects of putative IMBHs in the core
of dSphs on the very integrity of cold, long-lived substructure 
as that observed on the northeast side of the major axis of UMi. 
Although UMi has long been suspected of experiencing ongoing tidal disruption, 
regions with enhanced volume density and cold kinematics cannot be the result
of tidal interactions (Kleyna et al.~2003, hereafter K03; 
Read et al.~2006; S\'anchez-Salcedo \& Lora 2007).
This suggests that the secondary peak in UMi is a long-lived structure, 
surviving in phase-space because the underlying gravitational potential is 
close to harmonic (K03).
S\'anchez-Salcedo \& Lora (2007) derived an upper limit on the mass 
and abundance of massive dark objects in the halo of UMi
to avoid a quick destruction of the clump by
the continuous gravitational encounters with these objects.
In this work, we will assume that the dark halo is comprised of a smooth
distribution of elementary particles and then study the disintegration of
the clump,  placing contraints on the mass of a possible central IMBH in UMi.

\section{Initial conditions and $N-$body simulations}

\subsection{Ursa Minor and its clump}
\label{sec:genUMi}
Ursa Minor, located at a galactocentric distance of $R_{gc}=76\pm 4$~kpc 
(Carrera et al.~2002; Bellazzini et al.~2002), 
is one of the most dark matter-dominated dSphs in the
Local Group, with a central mass-to-light ratio $M/L\gtrsim 100 $ 
$M_{\odot}$/$L_\odot$ (e.g., Wilkinson et al.~2004). 
The measured central velocity dispersion is $17\pm 4$ km s$^{-1}$ (Mu\~noz
et al.~2005) and the core radius of the stellar component along the
semimajor axis is $\sim 300$~pc (Palma et al.~2003). 
UMi reveals several morphological peculiarities: 
(1) The shape of the inner isodensity contours of 
the surface density of stars appears 
to have a large ellipticity of $0.54$,
(2) the highest density of stars is not found at the center of 
symmetry of the outer isodensity contours but instead is offset southwest 
of center, (3) the secondary density peak on the northeast of the major
axis is kinematically cold. 

 \begin{figure*}
\plotone{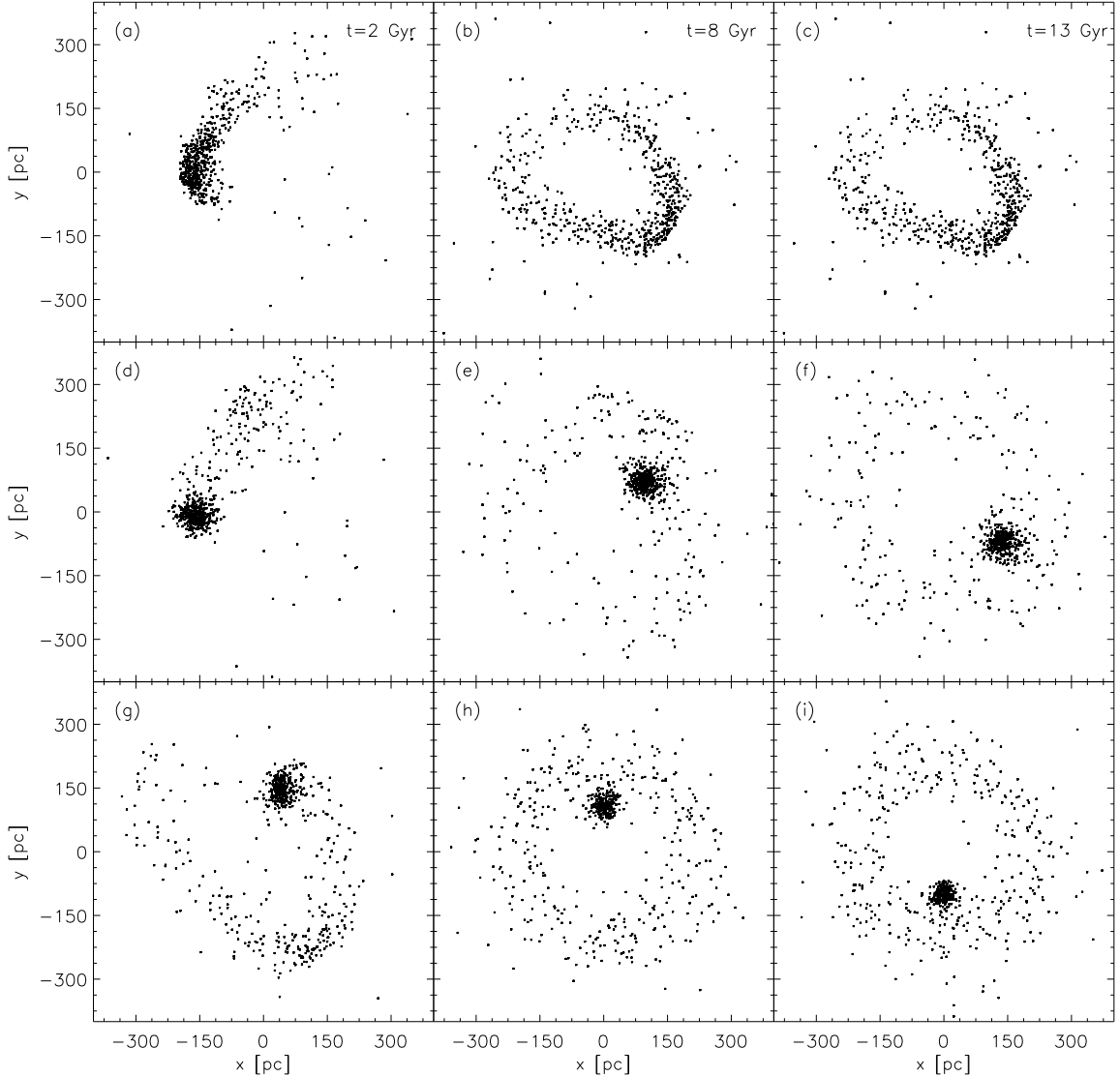}
 \caption{Snapshots at $t=2,$ $8$ and $13$ Gyr 
for $R_{1/2}=50$~pc and $R_{\rm core}=510$~pc without self-gravity
(top panel) and with self-gravity (middle panel). For comparison,
the snapshots for $R_{\rm core}=300$ pc and with self-gravity are 
also shown (bottom panel).}
 \label{fig:b510}
 \end{figure*}

The secondary density peak has a $1\sigma$ radius of $\simeq 1.6'$ 
($\sim 35$ pc at a distance of $76$ kpc) when
fitted with a Gaussian profile. 
The stellar distribution that forms this density excess is elongated not 
along the major axis of the isodensity contours of
the elliptized King model, but along a line at an intermediate angle between
the major and minor axes of these contours (Palma et al.~2003).
The bend in the isodensity contours indicates
that such clump is gravitationally unbound.
Interestingly, K03 found that the velocity of stars within 
a $6'$ ($130$ pc) radius aperture on the clump are best fitted by
a two-Gaussian population, one representing the underlying $8.8$ km
s$^{-1}$ population, and the other with a line-of-sight velocity dispersion of
$0.5$~km~s$^{-1}$. 
Although the value of the cold population is ill determined, 
we can be certain that the velocity dispersion is 
$<2.5$~km~s$^{-1}$ at a 95\% confidence level.
The mean velocity of the cold population is equal to the systemic
velocity of UMi, implying that either the orbit is radial and the
clump is now at apocenter or the orbit lies in the plane of the sky (K03).

\subsection{Initial conditions}
We consider the evolution of a clump inside a rigid halo of dark
matter with a density law: 
\begin{equation}
\rho(r)=\frac{\rho_{0}}{\left(1+\left[r/R_{\rm core}\right]^{2}\right)^{1/2}},
\label{eq:densitydm}
\end{equation}
where $\rho_{0}$ is the central density and $R_{\rm core}$ 
is the dark halo core radius. This profile was chosen in order to
facilitate comparison with K03. 
We explore different values for $R_{\rm core}$. 
Following K03, once $R_{\rm core}$ is fixed, we rescale the central
density to have a dark matter mass of $5\times 10^{7}$ $M_{\odot}$ 
inside $600$ pc,
which is approximately the maximum 
extent of the stellar distribution observed in UMi. 
Therefore, in all our models, we have a total mass-to-light ratio of  
$\left( M_{tot}/L_{V}\right) \approx 90 M_{\odot}/L_{\odot}^{V}$ 
inside $600$~pc, for a visual luminosity
$L_{V}=5.4\times 10^{5}$ $L_{\odot}$ (Palma et al.~2003).

For a normal stellar population with $M/L_{V}=2 M_{\odot}/L_{\odot}^{V}$, 
the stellar mass within the core radius of the dwarf is 
$\sim 4.5\times 10^{5} M_{\odot}$, whereas the dark matter mass is
$> 6.2\times 10^{6} M_{\odot}$. Therefore, the contribution to the
potential of the baryonic mass was ignored.

The initial density profile of the clump follows a Plummer mass distribution,
\begin{equation}
 \rho(r)=\frac{3}{4 \pi} \frac{M_{c} R_{p}^{2}}{(r^{2}+R_{p}^{2})^{5/2}},
\end{equation}
where $M_{c}=2\times 10^{4}$ $M_{\odot}$ is the total mass of the clump 
and $R_{p}$ is the Plummer radius. 
One should note that there is a simple relation for a Plummer model between 
$R_{p}$ and the half-mass radius: $R_{1/2}=1.3R_{p}$.
We use $R_{1/2}$ values 
between $25$ and $50$ pc to initialize our simulations. 
The clump's self-gravity is included to have a realistic and complete 
description of its internal dynamics because, as we will see later, 
the tidal radius of the clump may be larger than $R_{1/2}$  
when large values of $R_{\rm core}$ are used. 

The clump is dropped at the apogalactocentric distance of $200$ pc from the
UMi center with a certain tangential velocity $v_{T}$, which defines
the eccentricity $e$ of the orbit.
The orbit of the clump lies in the $x$-$y$ plane, which is also the plane 
of the 
sky.  For simulations with a central BH, it is clear that
the evolution of the clump depends on its orbital eccentricity.
For instance, we found that for a radial orbit, the BH 
dissolves the group of stars in its first passage through the galactic center. 
In order to provide an upper limit on the mass of the BH and
since there is more phase space available for nearly circular orbits 
than for radial ones,
we will take, in most of our simulations, a rather circular orbit
with $e=0.5$, which corresponds to $v_{T}=5.7$ km s$^{-1}$ for 
$R_{\rm core}=510$ pc.
However, to our surprise, the disintegration time was found to be
insensitive to the initial eccentricity of the clump's orbit as long
as $e<0.87$.
Altogether, there are essentially three model parameters that 
we have explored in our simulations: $R_{\rm core}$, $R_{1/2}$ and $M_{BH}$.

\subsection{$N$-body simulations}
We developed an N-body code that links an individual timestep to each particle
in the simulation. Only for the particle that has the minimum associated time,
the equations of motion are integrated (with a second order predictor-corrector 
method). This ``multi-timestep'' method 
reduces the typical CPU time of direct, particle-particle integrations 
($\propto N^{2}$, with $N$ the number of particles), 
and allows integrations of systems of $\sim$1000 particles to be 
carried out with relatively short CPU times. 

The cluster density is so tiny that the internal relaxation
timescale is very long ($\sim 11$ Gyr for an initial cluster with
$R_{1/2}=50$ pc). Therefore, although our code is suitable
to include two- and three-body encounters, the cluster behaves as
collisionless and internal processes such as evaporation do not
contribute to the dissolution of the cluster. For the same reason, all the
particles have the same mass, and the presence of binaries
was ignored. 

All the simulations presented in this paper used $600$ particles, each one
having a mass of $33 M_{\odot}$. 
We chose a smoothing length of $0.7$ pc, which is approximately
1/10th the typical separation among cluster particles within $R_{1/2}$ at
the beginning of the simulation. The convergence of the
results was tested by comparing runs with different softening length
and $N$.  The effect of adopting a dfferent smoothing radius between
$0.1$ and $1$ times the typical distance was found to be insignificant.
We run the same simulation with different $N$ ($N=200, 400, 600$ and
$1800$ particles), and convergence was found for $N\geq 400$.

In order to validate our simulations,
we checked that, when the cluster evolves at isolation, the Plummer
configuration is stationary and that the energy is conserved
over $12$ Gyr.
To be certain that the interaction with the BH is resolved well,
we compared the change in kinetic energy of the cluster,
when colliding with a particle of mass $10^{5} M_{\odot}$, moving at $200$
km s$^{-1}$, 
with the predictions in the impulse approximation (Binney \& Tremaine
2008), for different impact parameters, and found good agreement 
(differences less than $10\%$) between them. 

We were also able to reproduce K03 simulations 
of the evolution of an unbound clump (ignoring self-gravity) 
in the gravitational potential created by the mass distribution given by
Eq.~(\ref{eq:densitydm}).
We confirm the K03 claim that even if the clump is initially
very compact, with a $1\sigma$ radius of $12$ pc, 
cusped halos cannot explain the survival of the substructure
for more than $1$ Gyr, while the substructure can persist for $\sim 12$
Gyr in halos with core radii $2$--$3$ times the clump's orbit. 

\begin{figure*}
\plotone{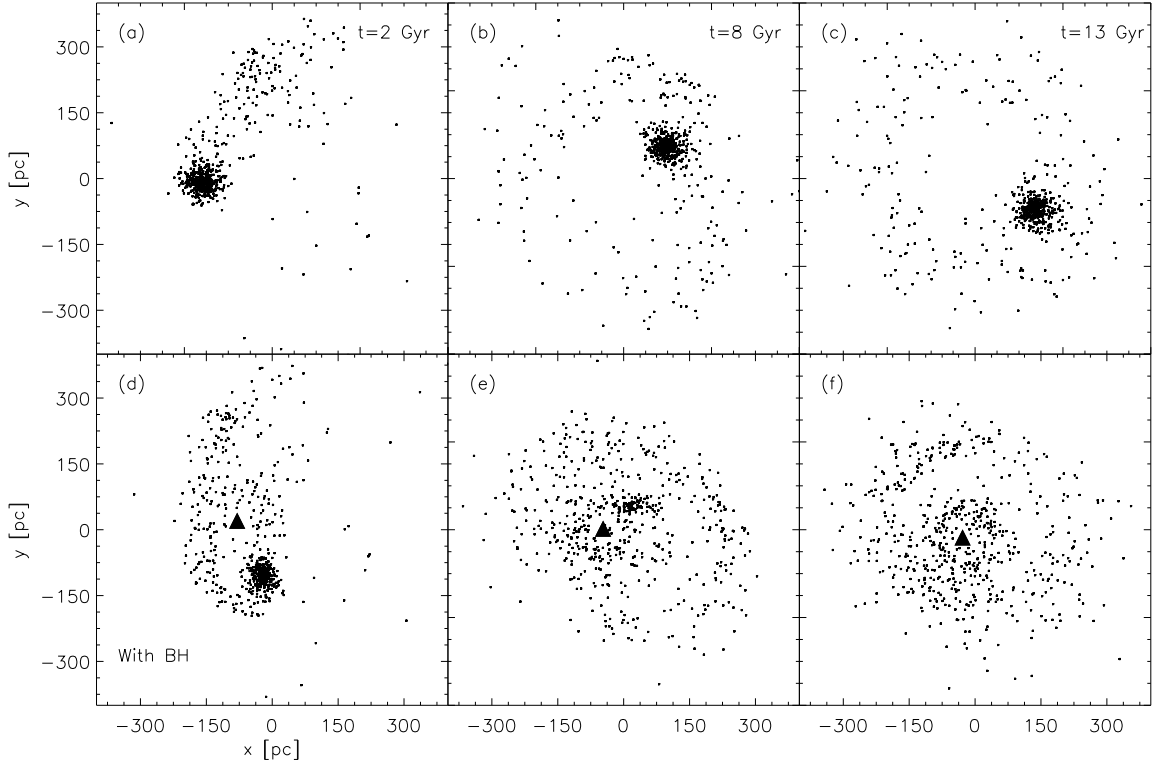}
 \caption{Snapshots at $2,$ $8$ and $13$ Gyr for a simulation
with $R_{1/2}=50$~pc and $R_{\rm core}=510$~pc (top panel), and 
the same simulation but  with a BH (triangle)
with mass $M_{BH}=3\times10^{4}$~$M_{\odot}$ (bottom panel).} 
 \label{fig:std1}
 \end{figure*}

\section{RESULTS}
\subsection{Simulations without BH: the role of self-gravity}
We have first examined the evolution of a clump with $R_{1/2}=50$ pc
embedded in a galaxy with a large core of $510$ pc (see Fig.~\ref{fig:b510}).
Note that when self-gravity is included, an initial $R_{1/2}$ value
of $\sim 15$ pc, as that used in K03, is no longer realistic because
the clump remains too compact over its lifetime to explain its present 
appearance.  With self-gravity, 
strong substructure continues to persist
for a Hubble time, while in the non-self-gravitating case 
the substructure is completely erased in $\sim 10$ Gyr. 
The evolution of the same 
self-gravitating clump but in a galaxy with 
$R_{\rm core}=300$ pc is also shown in Fig. \ref{fig:b510}. As expected,
when the size of the galaxy core is decreased, the clump 
does not survive so long, but we can have a galaxy core of $300$ pc
without causing the clump to disintegrate in $13$ Gyr.
We confirm the K03 claim that  
cored halos allow the structures to remain uncorrupted for a 
Hubble time. Moreover, we find that, when including self-gravity, 
a galaxy core of $1.5$ times the clump's orbit
is enough to preserve the integrity of the clump.

\subsection{Simulations with BH}
If UMi hosts an IMBH, the density substructure is erased not only by the orbital
phase mixing and the tidal field of the galaxy but also by the 
gravitational interaction with the hypothetical
BH. Since the destructive effects depend on the mass of the BH,
we can establish an upper limit on the mass of the BH in UMi by 
imposing that the BH must preserve the longevity of the clump for
more than $10$~Gyr.
We proceeded to add a massive particle of $3\times10^{4}~M_{\odot}$
at the center of the galaxy potential, emulating a BH. 
Figure~\ref{fig:std1} shows the evolution of a self-gravitating clump with an
initial size of $R_{1/2}=50$ pc in a galaxy with $R_{\rm core}=510$~pc.
Under the influence of the gravitational pull exerted by the clump, 
the BH, initially at rest, is displaced from the center of UMi. The
azimuthal orbital lag angle of the BH in the $(x,y)$ plane
is $\approx \pi/2$. 
As a consequence,
the clump feels a gravitational drag and loses angular momentum
which is transferred to the BH. Due to the angular
momentum loss, the clump starts to spiral into the center. At $2$ Gyr,
we can see that, in fact, the clump has smaller orbital radius
than it would have in the absence of the BH (see Fig.~\ref{fig:std1}).
The BH, on the other hand, spirals outwards to
larger radii until it reaches its maximum galactocentric radius. 
At $t=4$ Gyr, the clump reaches the galactic center and this 
``exchange'' of orbits starts all over again. 
Therefore, when looking at the BH of UMi, one should bear in mind
that the BH does not necessarily settle into the center of the
galaxy. It would be interesting to perform N-body simulations including
the stellar background to elucidate if the two observed off-centered regions
with the highest stellar density (e.g., Palma et al.~2003) are a 
consequence of the dynamical response to the BH. 
The minimum distance between the BH and the center of
mass of the clump is $\sim 100$ pc in this model.
Since the orbits of the clump and the BH never cross, the 
tidal disruption of the group of stars can be described as 
a secular process of stellar 
diffusion in phase-space. This slow relaxation process forms 
a stellar debris of stripped stars that move on a galactic orbit very
similar to that of the clump itself.

\begin{figure}
\plotone{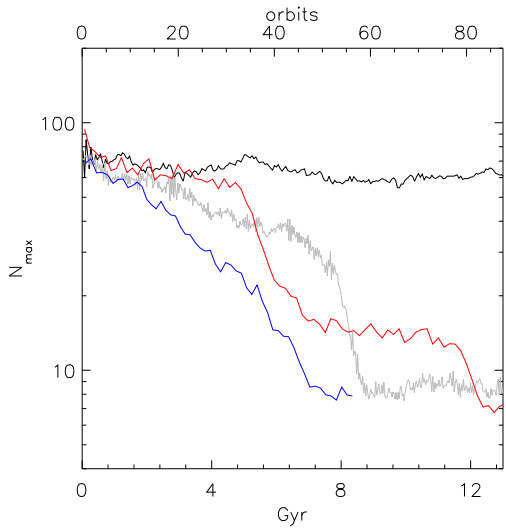}
 \caption{Smoothed curves of ${\mathcal{N}}_{max}$ as a function of time for
runs \# 0 (black line), \# 3 (grey line), \# 5 (red line) and \# 6 
(blue line).}
 \label{fig:nmax}
 \end{figure}

In Fig.~\ref{fig:std1} we see that, under the influence of the BH,
the clump is completely dissolved at $8$ Gyr. 
In order to quantify the clump's evolution, we calculated a map of the
surface density of particles in the $(x,y)$ plane at any given time $t$. 
We sample this two-dimensional map searching for the parcel 
(of $20\times 20$ pc size) 
that contains the highest number of stars ${\mathcal{N}}_{\rm max}(t)$.
This region is centered at the remnant of the clump.
The number of stars ${\mathcal{N}}_{\rm max}$ as a 
function of time is shown in Fig.~\ref{fig:nmax}. 
As expected, ${\mathcal{N}}_{\rm max}$ decreases 
as the simulation evolves due to the spatial dilution of the clump
caused by tidal heating.
Once ${\mathcal{N}}_{\rm max}$ drops a factor $2$, which
occurs in $\sim 6$ Gyr,
the tidal disruption process is accelerated and 
${\mathcal{N}}_{\rm max}$ dramatically declines at $\simeq 8$ Gyr
(around the orbit $55$), which can be taken as the disruption time,
denoted by $t_{d}$. 
The disruption of the clump was confirmed by visual inspection of the
simulations.

Further simulations show that 
even if the eccentricity of the orbit is similar to that of the
isodensity contours of the surface density of UMi background stars, 
the disruption time is essentially the same.

We carried out simulations with $R_{1/2}=50$ pc, $R_{\rm core}=510$ pc
and $e =0.5$, but with different $M_{BH}$.
We found that the substructure disintegrates in $11$ Gyr 
for a $10^{4}$ $M_{\odot}$ BH, and in $2$ Gyr for $M_{BH}=10^{5}$ $M_{\odot}$.
We also explored different combinations for $R_{\rm core}$ and $R_{1/2}$.
A reduction of the core of the galaxy leads to a more stringent upper
limit on the mass of the BH. However, one can increase the longevity
of the structure if a smaller value for $R_{1/2}$ is adopted.
For instance, we also find that $t_{d}\simeq 8$ Gyr for
$R_{\rm core}=200$ pc, $R_{1/2}=25$ pc and $M_{BH}=3\times 10^{4}$ $M_{\odot}$.

In the absence of any knowledge about the initial dynamical state of the
BH, the most natural assumption is that the center
of the galaxy was its presumed birth site\footnote{The BH candidate in UMi
reported by Maccarone et al.~(2005) is placed at 
about $7$ pc from the center.}.
For completeness, we have also carried out simulations for an off-centered BH.
A compilation of the models is given in Table \ref{tab:parameters}.
When the BH is on radial orbit,
$t_{d}\simeq 4$--$9$ Gyr, for $M_{BH}=3\times 10^{4} M_{\odot}$.
Only in very special circunstances --when the BH is not on a radial 
orbit, the orbits are coplanar and the azimuthal lag angle is $\pi$--, 
the disruption of the cluster is less efficient because
the separation between the clump and the BH is larger on average over
the simulation. 

We conclude that, if UMi's clumpiness is a primordial artifact, then
the survival of the secondary peak imposes  
an upper limit on the mass of the putative
BH of $M_{BH}=(2-3)\times10^{4}$ $M_{\odot}$, if the BH originally 
lurked at the center.  
The extrapolation of the $M_{BH}-\sigma$ relation for elliptical 
galaxies (Gultekin et al.~2009)
predicts a $1.0\pm^{5.0}_{0.9}\times 10^{4}$ $M_{\odot}$ 
BH for UMi. Therefore,
our constraint is still consistent with both the extrapolated value
and that inferred by Maccarone et al.~(2005).

\begin{table*}
\centering
\caption{Relevant parameters and destruction times.}
\medskip
\begin{tabular}{@{}ccccccccc@{}}
\hline	
Run &   $R_{\rm core}$ & $R_{1/2}$ & BH position$^{\ast}$ & BH velocity & $M_{BH}$  &   $t_{d}$\\
\#&    pc & pc & at $t=0$ [pc]  & at $t=0$ [km s$^{-1}$] & $M_{\odot}$ &    Gyr &\\
 \hline
 0&  510 & 50 & $--$ & $--$ & 0 & $>14$ &\\
 1&  510 & 50 &$0$ & $0$  & $5\times10^{3}$ &    $>14$ &\\
 2&  510& 50 &$0$ & $0$  & $1\times10^{4}$ &    11 &\\
 3&  510 & 50 & $0$ & $0$  & $3\times10^{4}$ &     8.3 &\\
 4&  510 & 50 & $0$ & $0$  & $1\times10^{5}$ &     1.7 &\\
 5&  200 & 25 & $0$ &$0$  & $3\times10^{4}$ &    7.8 &\\
 6&  510 & 50 & (0,-50,0) & 0 & $3\times 10^{4}$ & 5.5 & \\
 7&  510 & 50 & (0,0,-50) & 0 & $3\times 10^{4}$ & 8.5 & \\
 8&  510 & 50 & (0,-50,0) & 0 & $5\times 10^{4}$ & 3.5 & \\
 9&  510 & 50 & (0,0,-50) & 0 & $5\times 10^{4}$ & 5.0 & \\
 10&  510 & 50 & (0,0,-50) & 0 & $1\times 10^{5}$ & 1.8 & \\

\hline
\end{tabular}
\vskip 0.3cm
$^{\ast}$ The clump is initially at the position $(200,0,0)$.
\label{tab:parameters}
\end{table*}

\section{Discussion and conclusions} 
\label{conclusions}
\noindent

Our self-gravitating simulations confirm the claim of K03 that
the dark halo of UMi must have a core radius of $\sim 300$ pc, comparable
to the core radius of the underlying stellar population,
in order to preserve the integrity of this clump. 
While the firm dynamical detection of an IMBH in any dSph galaxy is 
challenging because it requires observations of the velocity dispersion
of stars deep into the core,
we have demonstrated that the very integrity of kinematically cold substructure
in dSph galaxies may impose useful limits not only on the core of
the galaxy but also on the mass of putative BHs.
In the case of UMi, the maximum mass of a BH initially seated
at the centre of the potential or initially on radial orbit, is 
$(2-3)\times 10^{4}$ $M_{\odot}$. 
When searching for direct detection of the possible IMBH in UMi (e.g.,
Maccarone et al.~2005), one should keep in mind that the BH may be offset 
from the
galactic center because of the gravitational pull exerted by the clump.

As pointed out by the referee, the contour map of the surface 
brightness of the nucleus of M31
also has a bright off-center source (Fig.~2 of Lauer et al.~1993),
which was interpreted not as a separate stellar system but as the 
apoapsis region of an eccentric stellar disk orbiting a central 
massive BH (Tremaine 1995; Lauer et al.~1996).
The timing problem of the clump may be circumvented if one assumes
that the secondary peak of UMi is the part of a ring close to apoapsis. 
Since the stars forming the secondary
peak in UMi have a mean velocity equal to the systemic velocity of UMi,
the ring should lie close to the plane of the sky and should be very
eccentric.
A serious difficulty with this scenario is that the apsides of the
orbits need to be extremely aligned. Our simulations have shown
that this alignment is not possible if the ring is the tidal debris
of a stellar cluster. 
We do not know a satisfactory mechanism to explain the
formation of a very eccentric stellar ring at scales of a galaxy with
the required apsidal alignment.

\acknowledgments
The thoughtful comments and suggestions by an anonymous referee 
have greatly improved the paper quality.
V.L, A.C.R \& A.E. thank financial support from
grant 61547 from CONACyT.
F.J.S.S.~acknowledges financial support from CONACyT
CB2006-60526 and PAPIIT IN114107 projects.
V.L.~wish to thank Stu group.


%
%

\end{document}